# Reinvestigation of high pressure polymorphism in Hafnium metal

K K Pandey[1*], Jyoti Gyanchandani[2], M Somayazulu[3], G K Dey[2], Surinder M Sharma[1] and S K Sikka[4]

*[1]High Pressure & Synchrotron Radiation Physics Division, Bhabha Atomic Research Centre, Mumbai-400 085, India*

*[2]Materials Science Division, Bhabha Atomic Research Centre, Mumbai-400 085, India*

*[3]Geophysical Laboratory, Carnegie Institution for Science, Washington, D.C. 20015, USA*

*[4]Indian National Science Academy, New Delhi-110 002, India*

*Corresponding author: K. K. Pandey, High Pressure & Synchrotron Radiation Physics Division, Bhabha Atomic Research Centre, Mumbai-400 085, India.
Email ID: kkpandey@barc.gov.in

# Reinvestigation of high pressure polymorphism in Hafnium metal

There has been a recent controversy about the high pressure polymorphism of Hafnium (Hf). Unlike, the earlier known α→ω structural transition at 38±8 GPa, Hrubiak et al (2012) did not observe it till 51 GPa. We have reinvestigated the room temperature phase diagram of Hf, employing x-ray diffraction (XRD) and DFT based first principles calculations. Experimental investigations have been carried out on several pure and impure Hf samples and also with different pressure transmitting media. Besides establishing the significant role of impurity levels on the high pressure phase diagram of Hf, our studies do not support the results of Hrubiak et al. (2012). The structural sequence, transition pressures, the lattice parameters, the c/a ratio and its variation with compression for the α and ω phases as predicted by our *ab-initio* scalar relativistic (SR) calculations are found to be in good agreement with our experimental results of pure Hf.

## I. INTRODUCTION

Hf belongs to Group IV B elements of the Periodic Table. Though the role played by *d*-orbitals in controlling the physical properties of Hf makes it interesting from basic science point of view, it is also an important element for usage in the nuclear industry. Hf and its alloys have high neutron absorption cross-section and are therefore used as neutron poison in the control rods of the nuclear reactors. Presence of ω-phase (occurrence of which under pressure is discussed in detail in this manuscript), embrittles these alloys and has deleterious effect on its mechanical properties.

At ambient conditions, Hf crystallizes in the hexagonal-close packed (hcp) structure, commonly referred to as the α-phase, like its homologues Ti and Zr. At higher temperatures, the hcp phase transforms to the body centred cubic (bcc) phase (β) at 2030 K, similar to Zr and Ti though the transition temperature in case of Zr and Ti are lower by a factor of ~2.[1] Subsequently, it was predicted that, like Zr and Ti, Hf should also follow the α→ω (a three atom hexagonal) →β transformation sequence.[2] However, unlike Zr and Ti, there are a very few high pressure experimental investigations on Hf. First high pressure experiments on Hf were done by Ming et al and they did not find any transition up to 39.5 GPa.[3] Soon after, Xia et al [4] compressed Hf in a diamond anvil cell (DAC) and showed, by using energy dispersive X-ray diffraction technique, that the α→ ω transition occurs at 38±8 GPa and the ω phase transforms to the β phase at 72±1 GPa. No other phase was observed up to 252 GPa. They related these two transitions to the previously observed two shock wave discontinuities by Bakanova et al [5] at 45 GPa (T=863K) and 60 GPa (T=1393K) respectively. More recently, Hrubiak et al have investigated the high PT phase diagram of Hf.[6] Their room temperature high pressure angle dispersive x-ray diffraction measurements did not show α→ ω transition up to 51 GPa. However, this transition was observed in compressed sample at elevated temperatures. This absence of high pressure phase transitions at ambient temperature is in disagreement with the results of Xia et al.[4] Hrubiak et al.[6] attributed this to the different pressure media (PTM) employed in the experiments. Xia et al did not use any PTM, while Hrubiak et al. used essentially NaCl. Earlier reports had established that α → ω transition pressure depends on the shear forces.[7] As these can arise in a DAC through non- hydrostatic environment, the dependence of the transition pressure on PTM is in principle understandable.[7-9] The amount of impurities present in the sample also affect the transition pressures. For example Vohra et al [10] reported that for Ti, change of oxygen content from 785 ppm to 3800 ppm shifted $P_{α→ω}$ from 3 GPa to above 8 GPa. In both the studies on Hf mentioned above [4,6], high purity samples were reported to be used. However, no chemical composition has been provided by the authors, except for a statement in ref. [4]

that their sample contained trace amounts of Zr.

This structural sequence in Hf i.e.; α → ω → β, has also received much attention theoretically. Table I summarizes the published theoretical results. As can be seen from the table, these *ab-initio* investigations show considerable scatter in the transition pressures.

These ambiguities in various experimental and theoretical investigations need to be addressed. With these motivations, we have revisited the high pressure polymorphism in Hf employing both experimental and theoretical methods. In disagreement with recent inference by Hrubiak et al, our investigations re-establish the room temperature α→ ω→ β transformation sequence in pure Hf where α→ω transition occurs above ~50.6 GPa and ω→β transition occurs above ~65 GPa. Computationally, the α→ω transition is found to be better represented by SR calculations and formation of the bcc (β) phase at higher pressures by both SR and SR+SO approximations.

## II. METHODOLOGY

### A. Experimental methods

High pressure x-ray diffraction measurements were carried out on three Hf samples with different purity levels viz.( a) 99.7 % pure Hf, other impurities: Ti and Zr <0.1%, Fe+Ni+Cr < 0.1 %, oxygen ~ 1600ppm, nitrogen ~ 25ppm (hereafter referred as Hf-a), (b) 95.6 % pure Hf, other impurities: Ti and Zr < 0.2 % , Fe+ Ni < 0.7 %, Co < 2.3 % , Mo < 1.22%   oxygen ~ 30 ppm, nitrogen ~ 30 ppm (hereafter referred as Hf-b),(c) 92.5 % pure Hf (hereafter referred as Hf-c). High pressure diffraction measurements on sample Hf-a were performed at Advanced Photon Source (USA) using λ=0.4066 Å up to 90.7 GPa. Samples Hf-b and Hf-c were studied up to 67 GPa and 73 GPa respectively in angle dispersive mode at beamline BL-11, located on a bending magnet port at Indus-2 synchrotron source, RRCAT, Indore (India) [16] using λ=0.7008 Å. Neon was used as pressure transmitting medium in the case of sample Hf-a whereas 4:1:: methanol: ethanol mixture was employed for samples Hf-b and Hf-c. Along with pressure transmitting medium, Cu/Au powder were also loaded inside the small sample chamber (ϕ ~100μm) prepared in tungsten/rhenium gaskets pre-indented to ~ 30μm for all the samples. Equation of state of these standard pressure calibrants (Cu/Au) were used for pressure estimation inside DAC.[17] 2D diffraction images recorded using MAR345 imaging plate detector were converted to 1D diffraction profiles using FIT2D software.[18,19] Lattice parameters and weight fraction of different phases of Hf were determined by Rietveld refinement as implemented in GSAS software.[20,21]

**B. Computational methods**

First principles structure optimizations were carried out through total energy calculations at 0 K using the full potential linearized augmented plane wave (FP-LAPW) method, as implemented in WIEN2K[22], employing the Perdew Burke Ernzerof generalized gradient approximation (GGA) for exchange and correlation effects. A systematic study in two sets of calculations was performed: the first one was using the LAPW method within scalar-relativistic (SR) approximation. In the second one, additional relativistic effects in the form of spin-orbit (SO) coupling were added using second variational method in the scalar relativistic basis orbitals. Here, local orbitals with $p_{1/2}$ radial basis were also used. Within each set, then, the computations have been carried out, both, with and without optimization of the *c/a* ratio, at each volume, for the hcp and the ω structures. Most of the earlier computations are carried out either with fixed c/a ratio or with c/a ratio optimized only at equilibrium volume. The ω structure is of $AlB_2$ type with space group P6/mmm, having 3 atoms in its unit cell at (0,0,0), (1/3, 2/3, 1/2) and (2/3,1/3,1/2).

The SR+SO calculations presented here on hcp, ω and bcc structures were carried out with a second variation energy cut off '$E_{cut}$'= 5 Ryd. 5d, 6s and 6p electrons were treated as valence states while the 4f, 5s and 5p electrons were treated as semi-core states. For sampling the Brillouin zone, a grid of 8000 k points was employed. The plane wave cutoff parameter $R_{MT} K_{MAX}$ was fixed at 9. The muffin tin radius $R_{MT}$ was set to 2.0 for the SR calculations but for SR+SO calculations it was set to 2.3 a.u. The self consistent cycle in each case was run till the energy convergence criterion of $10^{-5}$ Ryd was reached.

## III. RESULTS AND DISCUSSION

**A. Experimental**

Figure 1 shows stacked high pressure diffraction patterns of Hf-a sample at a few representative pressures. Rietveld analysis of the pattern at ambient conditions confirms the phase to be hexagonal close packed (α-phase) with lattice parameters, *a*= 3.1975 Å and *c/a* =1.5802, which give an equilibrium volume of 22.36 Å$^3$/atom and matches fairly well with earlier reported values.[4,6] This phase remains stable up to 50 GPa. At 50.6 GPa ($V/V_o$=0.74) new diffraction peaks start emerging corresponding to ω phase. The weight fraction of ω phase reaches 22.8 % at 58.5 GPa and 76.8 % at 62.1 GPa. However, even before the transformation to ω phase gets completed, β phase (bcc structure) emerges and the sample is found to be completely transformed to this phase at 67.4 GPa ($V/V_o$=0.68). This phase transition sequence is traced back on pressure release, however, with relatively larger hysteresis in α↔ω transformation as compared to ω↔β transformation. Relative abundances of different phases, reflecting the hysteresis are shown in figure 2.

These observations indicate slightly different transformation pressures as compared to those reported by Xia et al. where transformation to ω phase have been shown to start at 45.8 GPa ($V/V_o$=0.78) and complete by 58.3 GPa . Transformation to β phase was shown to be occurring at relatively higher pressures i.e. above 71.5 GPa ($V/V_o$=0.69) and complete by 78.4 GPa.

As mentioned above, Xia et al. performed high pressure experiments on Hf without any pressure transmitting medium while Hrubiak et al used NaCl, which is also not a good pressure transmitting medium.[23] The differences in observed transition pressures could be due to different hydrostatic environments, as suggested by Hrubiak et al too.

Though completely hydrostatic environment is illusive under high pressures above ~ 10 GPa, due to solidification of pressure transmitting media, still methanol: ethanol mixture or neon are well known to provide much better hydrostatic environment as compared to NaCl.[9,23,24] As purity of our Hf-a sample is comparable with the one reported by Hrubiak et al., and it shows the α→ω transition below 51 GPa with neon as the pressure transmitting medium whereas Hrubiak et al couldn't observe it even with relatively more non-hydrostatic pressure transmitting medium i.e NaCl, the observed discrepancies cannot be ascribed to pressure transmitting medium. Published P-T pathway of their experiment (figure 1 ref. 6) show that, at ambient temperature, they did not pressurize the sample at all beyond 51 GPa. Their high pressure measurements up to 67 GPa were performed only at elevated temperatures.

High pressure measurements on Hf-b sample exhibit the α→ω phase transition above ~ 56 GPa. Recorded x-ray diffraction patterns at a few representative pressures are shown in figure 3(a), (b). We could not get β-phase in Hf-b as the highest achieved pressure in these measurements was 67.1 GPa. On pressure release, this sample also shows large hysteresis in α↔ω transition. Despite different pressure transmitting medium and sample purity both the samples viz. Hf-a and Hf-b exhibit similar hysteresis loop of abundance (figure 4) implying a little role of these differences. It is interesting to note that the mean pressure value of α →ω transition is nearly the same as that of Xia et al.[4] Sample Hf-c was compressed to a maximum pressure of 73 GPa. However, this sample did not show the α to ω transition up to this pressure.

High pressure compression of lattice parameters in α-phase is also found to be similar in case of Hf-a and Hf-b samples (figure 5). The *c/a* ratio of α-phase is found to increase from ~ 1.58 at ambient pressure to 1.625 at ~ 55 GPa before its transformation to ω-phase. Our theoretical results also show an increase in *c/a* ratio with increasing pressure, that too with a remarkable similarity to the experimental observations (figure 5, figure 8(a) for the hcp case). Therefore, disagreeing with earlier contradictory reports (references given in ref. 6), we feel and find that this increase in *c/a* ratio is not an artifact arising from the non-hydrostatic stresses in a DAC.

Figure 6 gives our experimentally determined pressure - volume changes in Hf-a sample. Third order Birch Murnaghan equation of state [25] fitted to obtain the equation of state (EOS) parameters for the hcp phase gives the $B_o$(Bulk modulus) as 116.74 GPa and $B_o^{'}$(pressure derivative of $B_o$ ) as 3.0423 at the observed ambient experimental volume of 22.36 Å$^3$/atom . Corresponding EOS values as obtained by Hrubiak et al[6] at 22.30 Å$^3$/atom are 112.9 GPa and 3.29 respectively. Our corresponding theoretical values [26] at 22.45 Å$^3$/atom are 108 GPa and 3.42, respectively. It may be noted here that GGA overestimates the equilibrium volume slightly and when $V_0$ is large, correspondingly $B_0$ is small and vice versa.

Our results suggest that, in terms of α→ω phase transition, the high pressure behavior of Hafnium remains almost the same, up to the impurities' level of ~ 4.5 %. The small difference between the transition pressures may also partly be due the use of neon and ethanol-methanol pressure media in two cases. However, at still higher impurities concentration (> 7.5%), the behavior is different as in this study sample did not show any phase change up to 73 GPa.

**B. Computational**

Figure 7(a) and 7(b) shows the energy versus volume plot of the three structures for SR and SR+SO approximations, respectively. Table II summarizes the results of these computations with respect to phase transitions. As can be seen, the α→ ω transition occurs around 49 to 51 GPa at 16.68 Å$^3$/atom (figure 7(a)) when *c/a* is allowed to optimize along with volume (in hcp and ω structures). These values are estimated from enthalpy 'E+PV' versus pressure plot (not shown here). With *c/a* fixed this transformation takes place at a slightly lower pressure. In the SR+SO case, however, we do not observe the α→ ω phase transition, as energy curve for the ω phase stays above the one for hcp structure at all the pressures. For example, at 16.95 Å$^3$/atom i.e., the theoretical volume for the α→ ω transition (for the fixed *c/a* case of SR computations), we find the ω phase energy to be larger than that of the hcp phase by ~0.8 and ~0.5 mRyd for *c/a* optimized and *c/a* fixed cases respectively. The transition to the bcc (β) phase occurs around 67 GPa, at slightly lower pressure than in the SR case. (table II)

The computations in the SR+SO case were also carried out using $E_{cut}$ = 8 Ryd to see if it changes the relative position of hcp and omega curves near the expected region of transition. But we find that though the overall total energy of both the structures gets lowered by ~1mRyd for $E_{cut}$ = 8 compared to $E_{cut}$ = 5, nevertheless, the relative positions of the hcp and ω curves remains the same.

The *c/a* ratio for the α-phase has been found to increase monotonically with pressure in both the cases, i.e. in SR and SR+SO approximations and exhibit remarkable similarity with the experimentally obtained values (figure 8(a)). The qualitative behavior of *c/a* ratio in case of ω phase is also found to be similar to the experiments i. e. first increasing and then decreasing beyond a certain pressure. However, as shown in figure 8(b), the calculated pressure range of the crossover is found to be broader than that of the experiments.

As mentioned above, our calculations do not show any α → ω transition in SR+SO calculations. Such differences have been noticed in other studies also. For example, Hermann et al.[27] noted that Uuq (Z=114) prefers to crystallize in fcc structure in SR calculations while it prefers the hcp structure in SR+SO calculations. At high pressures, it, (i.e., Uuq) follows the structural sequence as fcc-hcp-bcc, very similar to its lighter homologue Pb. However, in SR+SO case, the hcp phase continues to remain stable implying no pressure induced transition. Again, in polonium the simple cubic phase exists because of strong spin-orbit interaction.[28] For Hf, we have investigated the reason for this discrepancy between SR and SR+SO calculations. It has been argued earlier by us[2,29], while analysing structural transitions in Group IVB elements, that structural phase changes occur due to the competition between the band structure and Ewald (Madelung) contributions in the total energy. We find in our computations that, at the same volume, the interstitial charge (difference between nuclear charge and charge inside the muffin tin radius) is higher in SR+SO case in comparison to the SR case, both in α and ω structures. This implies that with enhanced relativistic effects there is more 's' electron density in the interstitial regions.[30] This observation is similar to that of Fang et al[14] employing pseudopotential DFT calculations. Now, larger interstitial charge in SR+SO case, leads to a larger Madelung contibution.[30,31] This will favour hcp structure over ω structure (as Madelung constant for hcp is 1.79168 compared to 1.78856 for ω). Evaluation of the Madelung energy differences show that at a pressure of ~ 45 GPa (i.e. close to the transition pressure), the hcp should be more stable than the ω structure by about 5 mRyd in the SR+SO calculations compared to the SR. It may be pointed that lowering of the total energy curve of ω phase (SR+SO) by 1 mRyd leads to α → ω transition at about 45 GPa.

**IV. CONCLUSIONS**

Our high pressure investigations re-establish the existence of α→ω→β transformation sequence in Hafnium under room temperature conditions and, hence, we do not support the recent result of Hrubiak et al.[6] At room temperature, α↔ω transition exhibits large hysteresis encompassing a pressure range of over 30 GPa, similar to that observed by Xia et al.[4] Pressure transmitting media do not affect α↔ω transition drastically so long as the pressure is quasi-hydrostatic. However, sample impurity change the phase transformation pressure, as impurity concentration of ~ 7.5 % can even suppress the transformation to ω phase. Our *ab-initio* calculations in scalar relativistic (SR) approximation reproduce the experimental structural sequence in pure Hf. However, incorporation of spin-orbit (SO) interaction suppresses α→ω transition. An explanation is offered for this difference in terms of differences in s-electron density in the interstitial regions, which favor the more close packed structure. In both the approximations, optimizing *c/a* ratio produces transformation pressures closer to the experimentally obtained values.

**Table I.** Theoretically estimated transition pressures (GPa) for α → ω and ω→β transitions, as reported by different investigators.

| Authors | P(α → ω) | P(ω→β) | Method |
|---|---|---|---|
| Ahuja et al [11] | 13.9 | 30.7 | FP-LMTO (LDA) |
| Ostanin et al [12] | 32 | 140 | FP-LMTO (GGA) |
| Jomard et al [13] | | | FP-LMTO |
| | 12.7 | 47.6 | LDA |
| | 43.5 | 62.6 | GGA |
| Fang et al [14] | | | Pseudo-potential (VASP) |
| | 10.6 | 60.6 | LDA |
| | 12.2 | 56.5 | LDA+ spin orbit (SO) |
| | 24.5 | 78 | GGA |
| | 32.5 | 77.5 | GGA+ SO |
| Hao et al [15] | | | Pseudo-potential (VASP) |
| | 44.5 | 66.2 | GGA |

**Table II.** Calculated equilibrium volume at zero pressure ($V_0$), *c/a* ratio, volume V, $V/V_0$ and pressure corresponding to α→ω and ω→β transitions as deduced from SR and SR+SO calculations.

| | $V_0$ (Å³/atom) | *c/a* (hcp) | V ($V/V_0$) α→ω | P α→ω | V ($V/V_0$) ω→β | P ω→β |
|---|---|---|---|---|---|---|
| SR Optimize *c/a* | 22.554 | 1.582 | 16.68 (0.74) | 50 | 15.60 (0.69) | 69 |
| SR Fixed *c/a* | 22.552 | 1.582 | 16.95 (0.75) | 47 | 15.62 (0.69) | 70 |
| SR+SO Optimize *c/a* | 22.456 | 1.5813 | | | 15.72 (0.70) | 67* |
| SR+SO Fixed *c/a* | 22.453 | 1.582 | | | 15.86 (0.71) | 65* |

\* The transformation, here, is to the bcc (β) phase, though not directly from the ω phase

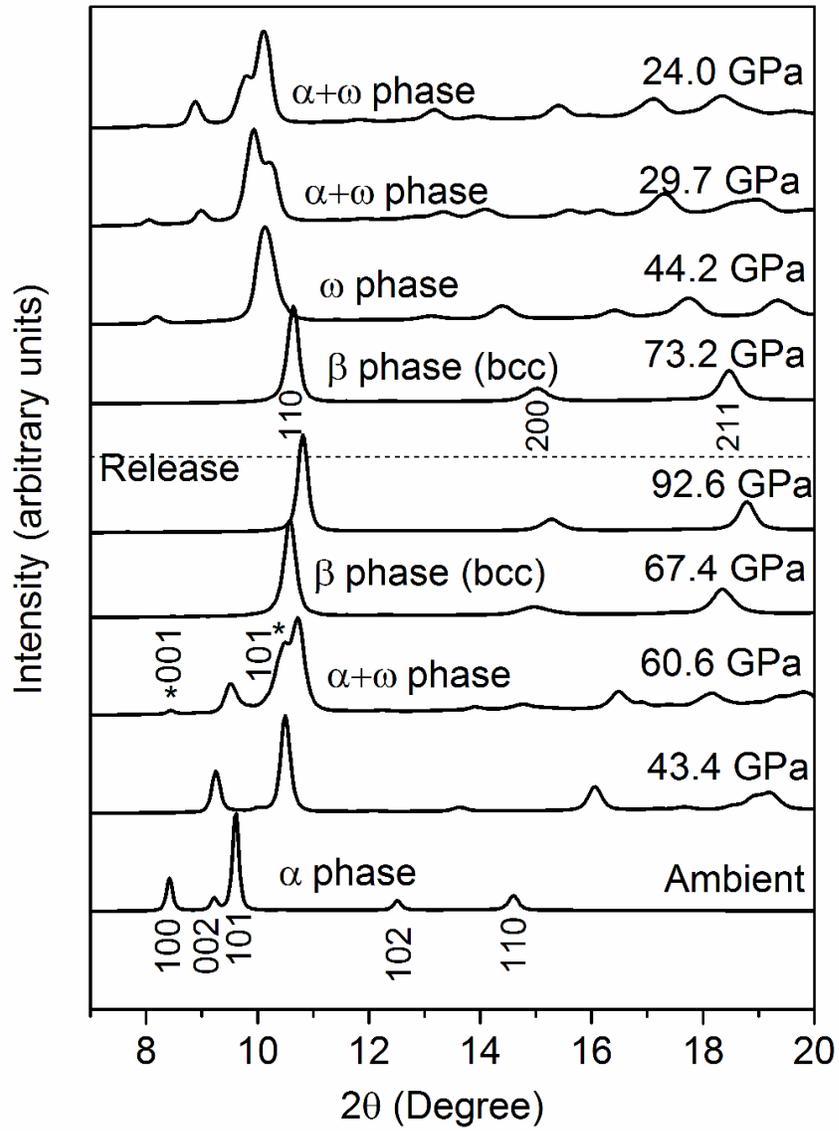

**FIG.1.** Stacked diffraction patterns of Hf-a at a few representative pressures.

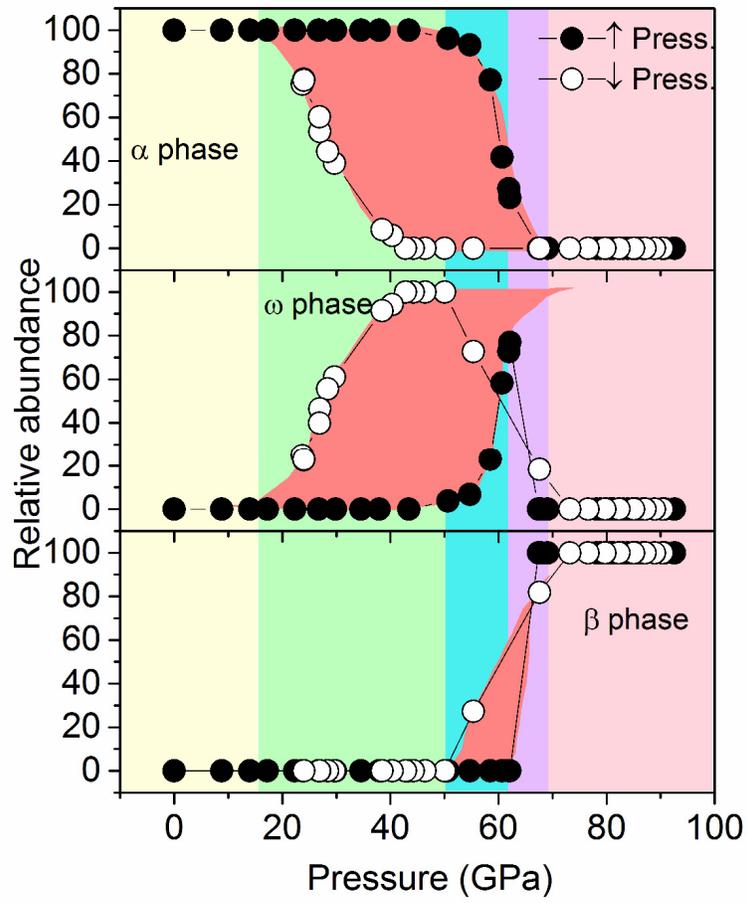

**FIG.2.** Percentage abundance of various phases of Hf-a sample as a function of pressure

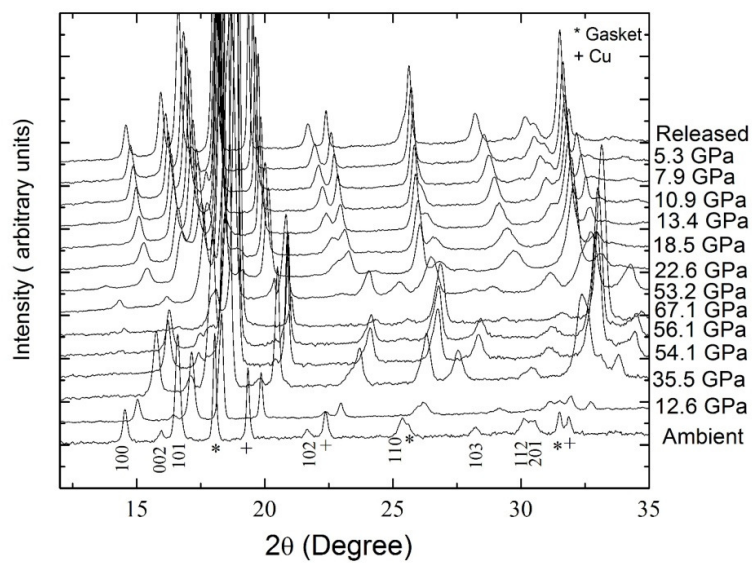

**(a)**

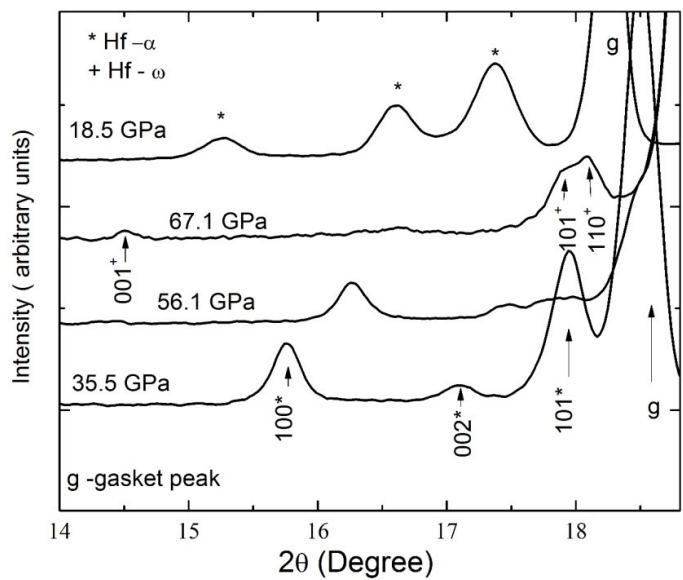

**(b)**

**FIG.3.** (a) X-ray diffraction patterns at a few representative pressures for the sample Hf-b. (b) Some of the chosen diffraction patterns at relevant pressures and up to the relevant degrees of angle for depicting diffraction peaks of α and ω phases for Hf-b.

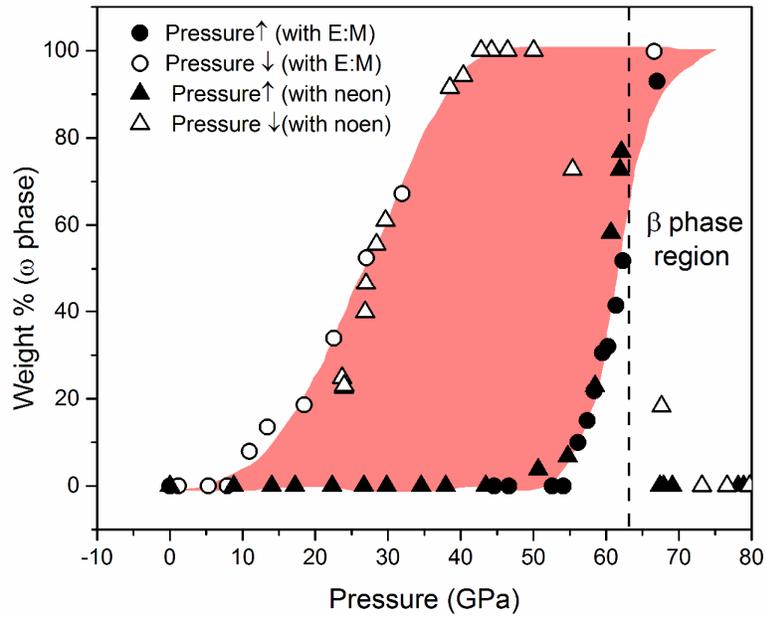

**FIG.4.** Percentage weight fraction of ω phase as a function of pressure in two separate measurements with different pressure transmitting medium and sample purity (sample Hf-a and Hf-b)

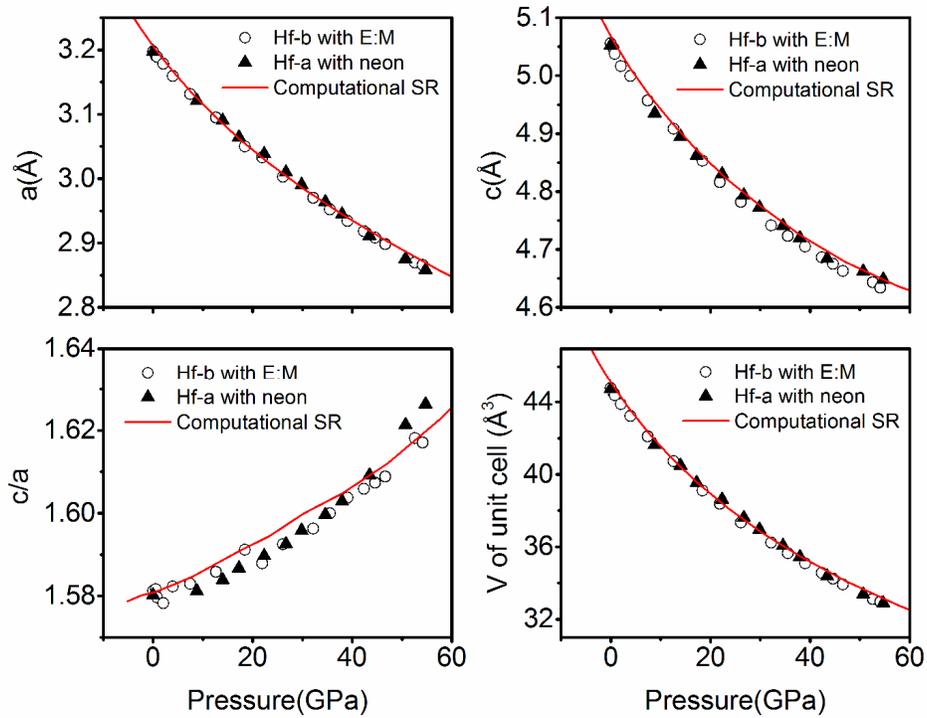

**FIG.5.** Observed variation in lattice parameters and unit cell volume as a function of pressure for α-phase of Hf with different pressure transmitting media. Also, plotted are the corresponding lattice parameters and unit cell volumes as estimated from our present SR computations.

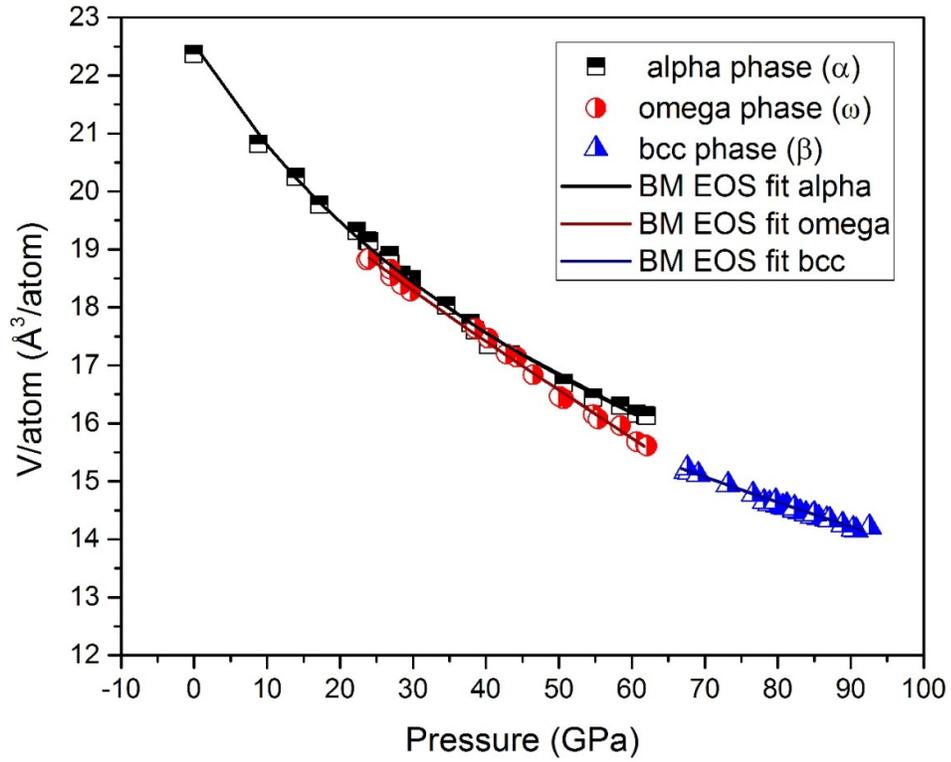

**FIG.6.** Experimentally determined pressure - volume changes in Hf-a sample.

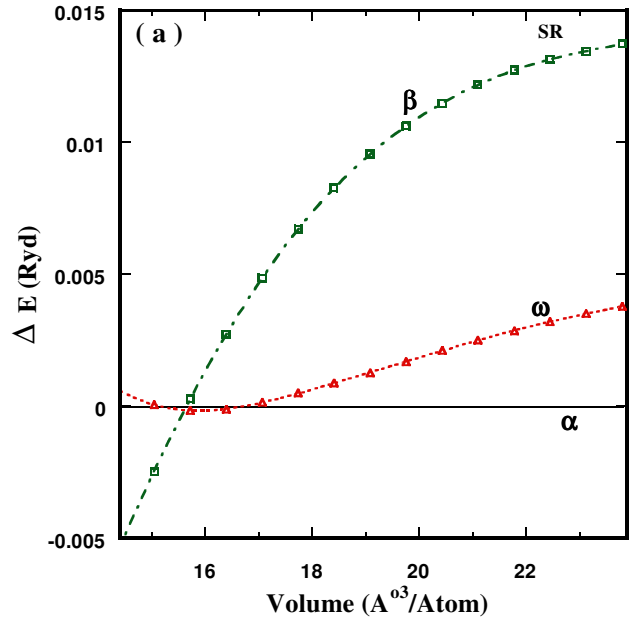

(a)

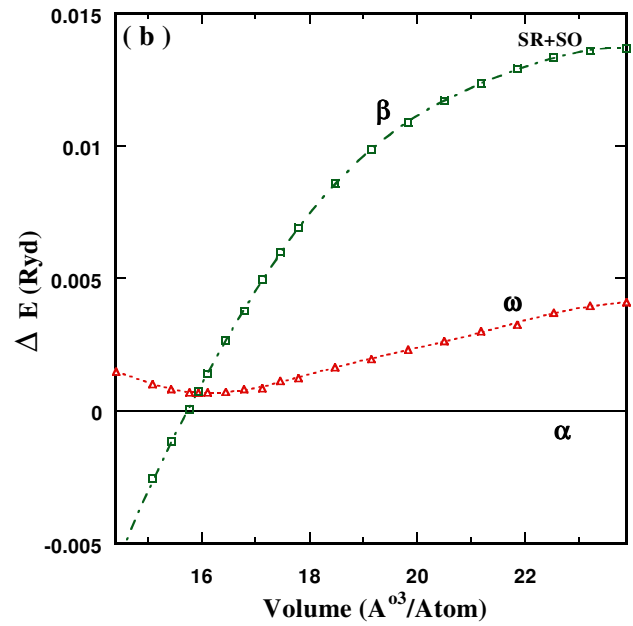

(b)

**FIG.7.** Energy differences of the bcc and ω structures with respect to the hcp structure versus Volume/atom for Hf using (a) SR and (b) SR+SO approximations.

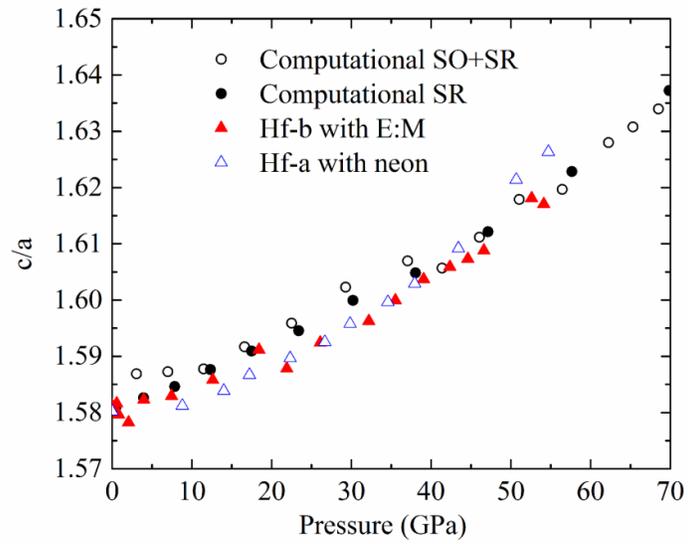

(a)

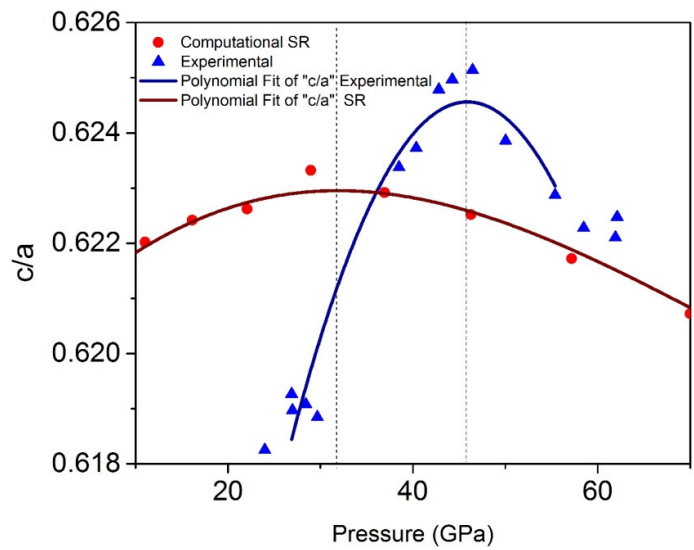

(b)

**FIG.8.** Experimental and theoretical *c/a* ratio of (a) α-phase (b) ω phase.